\documentclass[aps,pra,amsmath,notitlepage,twocolumn,superscriptaddress,nolongbibliography]{revtex4-2}
\pdfoutput=1
\usepackage[utf8]{inputenc}
\usepackage[english]{babel}
\usepackage[T1]{fontenc}
\usepackage{microtype}

\usepackage{amsmath,amsfonts,amssymb,amsthm,bm,bbm,bbold}
\usepackage{braket}
\usepackage{mathtools}

\usepackage[dvipsnames]{xcolor}
\usepackage[breaklinks=true,colorlinks=true,linkcolor=teal,urlcolor=teal,citecolor=teal]{hyperref}

\DeclareMathOperator{\Tr}{Tr}

\newcommand{\I}{\mathrm{i}}
\newcommand{\id}{\mathbb{1}}

\usepackage[normalem]{ulem}
\newcommand{\away}[1]{\textcolor{red}{\ifmmode\text{\sout{\ensuremath{#1}}}\else\sout{#1}\fi}}

\begin{document}

\title{Invasiveness of non-equilibrium quantum thermometry}

\author{Francesco Albarelli}
\email{francesco.albarelli@gmail.com}
\affiliation{Dipartimento di Fisica ``Aldo Pontremoli'', Università degli Studi di Milano, via Celoria 16, 20133 Milan, Italy}
\affiliation{Istituto Nazionale di Fisica Nucleare, Sezione di Milano, via Celoria 16, 20133 Milan, Italy}

\author{Matteo G. A. Paris }
\affiliation{Dipartimento di Fisica ``Aldo Pontremoli'', Università degli Studi di Milano, via Celoria 16, 20133 Milan, Italy}
\affiliation{Istituto Nazionale di Fisica Nucleare, Sezione di Milano, via Celoria 16, 20133 Milan, Italy}

\author{Bassano Vacchini}
\affiliation{Dipartimento di Fisica ``Aldo Pontremoli'', Università degli Studi di Milano, via Celoria 16, 20133 Milan, Italy}
\affiliation{Istituto Nazionale di Fisica Nucleare, Sezione di Milano, via Celoria 16, 20133 Milan, Italy}

\author{Andrea Smirne}
\affiliation{Dipartimento di Fisica ``Aldo Pontremoli'', Università degli Studi di Milano, via Celoria 16, 20133 Milan, Italy}
\affiliation{Istituto Nazionale di Fisica Nucleare, Sezione di Milano, via Celoria 16, 20133 Milan, Italy}

\begin{abstract}
One of the main advantages expected from using quantum probes as thermometers is non invasiveness, i.e., a negligible perturbation to the thermal sample.
However, invasiveness is rarely  investigated explicitly.
Here, focusing on a pure-dephasing spin probe in a bosonic sample, we show that there is a non-trivial relation between the information on the temperature gained by a quantum probe and the heat absorbed by the sample due to the interaction.
We show that optimizing over the probing time, i.e. considering a time-optimal probing scheme, also has the benefit of limiting the heat absorbed by the sample in each shot of the experiment.
For such time-optimal protocols, we show that it is advantageous to have very strong probe-sample coupling, since in this regime the accuracy increases linearly with the coupling strength, while the amount of heat per shot saturates to a finite value.
Since in pure-dephasing models the absorbed heat corresponds to the external work needed to couple and decouple the probe and the sample, our results also represent a first step towards the analysis of the thermodynamic and energetic cost of quantum thermometry.
\end{abstract}

\maketitle

\emph{Introduction.}---Estimating the temperature of a quantum system is a task of fundamental and practical importance.
Many quantum technologies require very low temperatures to exploit fragile nonclassical features, thus temperature must be assessed with great accuracy while disturbing the system as little as possible.
This is precisely the goal of quantum thermometry, a fertile research field at the intersection of quantum metrology, quantum thermodynamics and open quantum systems~\cite{DePasquale2018,Mehboudi2019}.
The accuracy of equilibrium quantum thermometry has been extensively discussed~\cite{Correa2015,Paris2016,Campbell2018a,Potts2019,Mehboudi2022a}. 
Going beyond equilibrium, temperature can also be estimated via quantum probes interacting with a thermal sample, commonly studied by modelling the probe as an open quantum system~\cite{breuer2002theory} and the sample as a bosonic~\cite{Brunelli2012,Correa2015,Kiilerich2018b,Cavina2018,Razavian2018,Khan2022,Oghittu2022} or fermionic~\cite{Mitchison2020,Oghittu2022,Mihailescu2022} environment, or by means of collisional approaches~\cite{Campbell2021a,Seah2019b,Shu2020a,OConnor2021a}.
Similarly, open quantum systems may also be used as probes to estimate other environmental parameters~\cite{Benedetti2014,Rossi2015,Bina2018,Tamascelli2020} and for quantum noise spectroscopy~\cite{Degen2016,Szankowski2017b,Wang2021a}.

In this Letter, we investigate the perturbation induced on the initial thermal state of the sample by the interaction with the probe.
We will call \textit{invasiveness} this feature of non-equilibrium thermometry protocols~\footnote{To avoid confusion, we stress that this is not the notion of invasiveness appearing in the context of Legget-Garg inequalities (which has also been connected to quantum metrology~\cite{Moreira2017}).}.
Having this goal, it is necessary to go beyond the standard paradigm of open quantum systems and consider also the dynamics of the environment, especially in the regime of strong coupling.Indeed, studying the dynamics of the environment is becoming crucial~\cite{Ptaszynski2019,Gribben2019,Brenes2020,Tamascelli2020a,Popovic2021a,Chen2021,Ptaszynski2022,Gribben2022}, especially in the context of strong-coupling quantum thermodynamics~\cite{Wiedmann2020a,Landi2021a,Talkner2020a}.
Moreover, since it is customary to model the system-environment dynamics as purely Hamiltonian, it is unclear if the sample will thermalize again after interacting with the probe.
Interestingly, in the continuum limit the thermalization of probe and sample may actually arise from purely Hamiltonian dynamics~\cite{Trushechkin2022a}.

Concretely, we propose to quantify the invasiveness of probe-based thermometric protocols in terms of the average heat absorbed by the sample, a choice informed by quantum thermodynamics.
We consider the spin-boson model \cite{Leggett1987}, where an environment of harmonic oscillators constitutes a thermal sample coupled (possibly strongly) to a spin probe.
In particular we focus on a coupling that preserves the probe's internal energy, inducing a pure dephasing dynamics that can be exploited for thermometry~\cite{Razavian2018,Gebbia2019,Candeloro2021,Wang2021a}.
Since the probe does not dissipate energy, one may think that the thermodynamic features of the model may be trivial.
However, external work is needed to couple and decouple the probe and the sample so that heat is dissipated into the environment~\cite{Marcantoni2017,Popovic2021,Francica2021}, perturbing the sample from its initial state of thermal equilibrium.

\emph{Dephasing dynamics of the probe.}---We consider a finite-dimensional probe system, with a generic Hamiltonian $H_S = \sum_j \epsilon_j | j \rangle \langle j |$, where $\ket{j}$ is the energy eigenbasis.
The environment, i.e. the sample, is modeled as an ensemble of noninteracting harmonic oscillators with free Hamiltonian $H_E = \sum_k \omega_k b^\dag_k b_k$.
System and environment are coupled by the interaction Hamiltonian $H_{I} = A_S \otimes \left( \sum_k  f_k b^\dag_k  + f_k^* b_k \right)$, with $A_S = \sum_j g_j | j \rangle \langle j | $.
The joint system-environment state evolves unitarily as $\rho_{SE} (t)= U(t) \rho_{SE} (0) U(t)^\dag$ with $U(t) = \exp \left[ -\I t \left( H_S + H_I + H_E \right) \right]$.
Since $[H_S, A_S] = 0$ the system undergoes a pure dephasing dynamics.
The populations of the energy levels are constants of motion, while the off-diagonal elements of the reduced density matrix $\rho_S(t) = \Tr_E \left[ \rho_{SE} (t)  \right]$ in the energy eigenbasis evolve as $\rho_S(t) = \sum_{ij} \rho_{S,ij}(0) e^{ - \left[ \Delta_{ij}(t) +\I \varphi_{ij}(t) \right] } |i \rangle \langle j |$, for appropriate real dephasing functions $\Delta_{ij}(t)$ and phases $\varphi_{ij}(t)$, see Appendix~\ref{app:puredephasing}.

We further assume an initial factorized state $\rho_{SE} (0) = \rho_S(0) \otimes \rho_E(0)$ and that the environment starts in a Gibbs thermal state $\rho_E(0) = {e^{-\frac{H_E}{T}}}/Z_T$, where $Z_T=\Tr[ e^{-\frac{H_E}{T}} ]$ is the partition function.
We choose units such that $\hbar = 1 $ and $\kappa_B=1$, so that both temperature and energy are measured as frequencies.
The dephasing functions $\Delta_{ij}(t) = ( g_i - g_j )^2 \Delta_T(t)$ are temperature-dependent:
\begin{equation}
  \label{eq:Gamma_ij_cont}
  \Delta_T(t) = \int_0^\infty \!\! d \omega \, J(\omega) \frac{1 - \cos \omega t}{\omega^2} \coth\left( \frac{\omega}{2 T} \right).
\end{equation}
We have also taken the continuum limit, informally $ \sum_k |f_k|^2 \mapsto \int_0^\infty d \omega J(\omega) $, where $J(\omega)$ is the spectral density that includes both the density of states of the sample and a non-uniform distribution of the coupling parameters $f_k$.
The phases $\varphi_{ij}(t)$ include both the free evolution due to $H_S$ and a contribution due to the interaction.
However, they do not depend on $T$ and are also irrelevant for energetic considerations, so we will neglect them (formally, working in a suitable rotating frame).

\emph{Heat absorbed by the sample.}---We define the average heat absorbed by the sample as the change in the expectation value of its Hamiltonian~\cite{Landi2021a} $Q(t) = \Tr_E \left[ H_E \left( \rho_E(t) - \rho_E(0)  \right) \right]$, with $\rho_E(t)= \Tr_S \rho_{SE}(t)$.
In the dephasing model we are considering, even if the system energy is preserved, the environment energy is not a conserved quantity since $[H_B,H_I]\neq 0$.
The absorbed heat can be obtained from solving the global dynamics (details in Appendix~\ref{app:puredephasing}, see also Ref.~\cite{Popovic2021a}); in the continuum limit it reads
\begin{equation}
  \label{eq:Qabs_cont}
  Q(t) = 2  \biggl( \sum_j \rho_{S,jj}(0) g_j^2 \biggr) \underbrace{\int_0^\infty d\omega J(\omega) \frac{1 - \cos \omega t}{\omega}.}_{ \textstyle  \equiv \mathbbm{Q}(t)} 
\end{equation}
Notice that for this pure dephasing model the heat is always positive: the environment always absorbs energy.
Moreover, it is independent of the temperature: the temporal dependence is completely determined by the ``bare'' spectral density.
In Eqs.~\eqref{eq:Gamma_ij_cont} and \eqref{eq:Qabs_cont} we have highlighted the quantities $\Delta_T(t)$ and $\mathbbm{Q}(t)$ encapsulating the time-dependence.
The initial state $\rho_S(0)$ only affects the absorbed heat as an overall multiplicative factor and when $g_j^2=g_k^2\,\, \forall j,k$, the heat $Q(t)$ is independent of $\rho_S(0)$.

For pure dephasing, the absorbed heat $Q(t)$ also corresponds exactly to the work needed to couple and decouple the system and the environment~\cite{Popovic2021,Francica2021}.
Since we are not modeling the coupling and decoupling explicitly, $Q(t)$ represents the work needed to perform an instantaneous coupling and decoupling, i.e. the parameters $g_j$ jump from 0 to their fixed value at time $0$ and the opposite at time $t$.
Thus, for probing schemes based on dephasing we are not only studying the invasiveness (i.e. heating of the sample) but also the work cost of thermometry (neglecting the cost of state preparation~\cite{Liuzzo-Scorpo2018,Lipka-Bartosik2018} and measurement~\cite{Deffner2016b,Guryanova2020}).

\emph{Two-level probe with Ohmic-like spectral density.}---
For the sake of concreteness, from now on we focus on a two-level probe, coupled through the operator $A_S = \lambda \sigma_z$, where $\lambda$ is an adimensional interaction-strength parameter, i.e. $g_0 = - g_1 = \lambda$ in Eq.~\eqref{eq:Gamma_ij_cont}.
The dephasing function reads $\Delta_{01}(t) = \Delta_{10}(t) = 4 \lambda^2 \Delta_T(t)$ and the absorbed heat $Q(t)=2 \lambda^2 \mathbbm{Q}(t)$.
We also focus on a spectral density of the form $J(\omega) = \omega \left( \omega / \omega_c \right)^{\!s-1} C(\omega,\omega_c)$, where $s$ is the so-called Ohmicity parameter and distinguishes three regimes: Ohmic for $s=1$, sub-Ohmic for $0<s<1$ and super-Ohmic for $s>1$.

In the main text we present results for an exponential cutoff $C(\omega,\omega_c)=e^{-\omega/{\omega_c}}$, which grants a closed-form expression for the dephasing function---originally derived in Ref.~\cite{Razavian2018} and reported in Appendix~\ref{app:dephasingOhm}---as well as a simple formula for the absorbed heat~\eqref{eq:Qabs_cont}
\begin{equation}
  \mathbbm{Q}^\mathrm{exp}(t) = \omega _c \Gamma (s) \left( 1- \frac{\cos \left[s \arctan\left( t \omega_c\right)\right]}{\left( t^2 \omega_c^2 +1\right)^{\frac{s}{2}}} \right),
  \label{eq:QabsExpCutoff}
\end{equation}
where $\Gamma (s)$ is the Gamma function.
In Appendix~\ref{app:other_spectral_densities} we show additional results for a Gaussian cutoff and a hard cutoff, for which the exchanged heat can be found analytically, see Appendix~\ref{app:dephasingOhm}, while the dephasing function~\eqref{eq:Gamma_ij_cont} is obtained by numerical integration.
While some phenomenology is different, the main qualitative features remain valid with different cutoffs.
\begin{figure*}[ht!]
  \centering
  \includegraphics{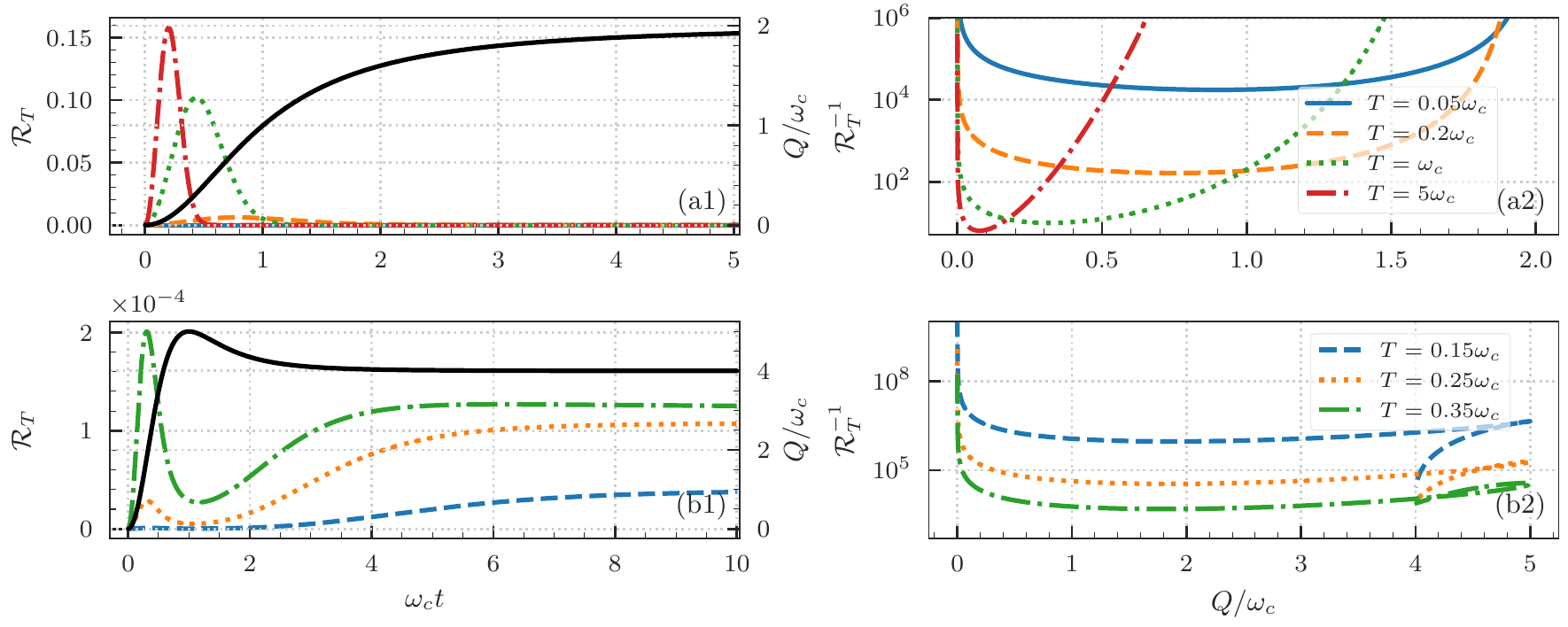}
  \caption{
  Top [panels (a)]: Ohmic spectral density $s=1$; 
  bottom [panels (b)]: super-Ohmic spectral density $s=3$; both for coupling $\lambda = 1$.
  Panels (a1) and (b1): absorbed heat (solid black line, units on the right) and quantum SNR (coloured lines for various temperatures, units on the left) as a function of time. 
  Panels (a2) and (b2): parametric plot of relative error for temperature estimation versus absorbed heat.
  Each line represents a different temperature, as shown in the legend.
  }  
\label{fig:QFIvsHeatPlot1}
\end{figure*}

\emph{Thermometric performance versus absorbed heat.}---Temperature is a parameter to be estimated from measurements on the probe.
The accuracy of the estimation is influenced by the measurement choice, formally a positive operator-valued measure (POVM), and by the classical estimator $\tilde{T}$ that turns the observed outcomes into a temperature estimate.
Since temperature is an energy scale parameter, it is common~\cite{Correa2015,Henao2021,Potts2019,Rubio2021}, and arguably more appropriate~\cite{Rubio2022}, to consider the relative estimation error.
Thus, we quantify the estimation accuracy with the signal-to-noise ratio (SNR), the inverse of the relative mean square error $\Delta^2 \tilde{T}/{T^2}$ of the estimator.

For unbiased estimators, the quantum Cramér-Rao bound (QCRB)~\cite{helstrom1976quantum,Braunstein1994} gives
\begin{equation}
  \frac{T^2}{\Delta^2 \tilde{T}} \leq M T^2 \mathcal{F}[\rho_T(t)] \equiv M \mathcal{R}_T(t) ,
\end{equation}
where $\mathcal{F}[\rho_T]$ is the quantum Fisher information (QFI) of the state $\rho_T$ with respect to the parameter $T$, expressed as $\mathcal{F}[\rho_T ] = \Tr \left[ \partial_T \rho_T L_T \right]$ with the hermitian symmetric logarithmic derivative operator $L_T$ defined by $2 \partial_T \rho_T = L_T \rho_T + \rho_T L_T $.
We have also introduced the dimensionless quantum signal-to-noise ratio (QSNR)~\cite{Paris2009} $\mathcal{R}_T= T^2 \mathcal{F}[\rho_T]$.
Here, $M$ is the number of identical shots of the experiment and this bound can be saturated asymptotically for large $M$ by choosing optimal measurements and estimators.
For a two-level probe in the initial state $\cos(\theta/2) \ket{0}+ \sin(\theta/2) \ket{1}$ the QFI for temperature estimation is $\mathcal{F}\left[ \rho_{T}(t) \right] = \left[ \sin \theta \, 4 \lambda^2 \, \partial_T \! \Delta_T(t) \right]^2 / \left( \exp\left[ 8 \lambda^2 \Delta_T(t)\right] - 1 \right)$
and it is attained by a projective measurement on $\sigma_x$ eigenstates~\cite{Razavian2018}; a balanced superposition $\theta = \pi /2 $ is optimal and will always be considered in what follows.

In Fig.~\ref{fig:QFIvsHeatPlot1} we plot the relative error and the absorbed heat as the probing time varies.
The parameter values for the plots are chosen to highlight a few of the different features that these figures of merit can display.
First of all, a certain amount of heat is inevitably absorbed by the sample, since this is due to the same interaction that imprints the information about the temperature on the probe.
As a consequence, during the initial part of the dynamics there is always a trade-off between the two quantities: to increase the accuracy we must allow the sample to absorb heat.
The absorbed heat is always positive for this model~\cite{Popovic2021}, but in general it has a nonmonotonic behaviour in time, as highlighted, e.g., by the curve in panel (b1) for $s=3$.
However, $Q(t)$ settles to a finite asymptotic value $\lim_{t \to \infty } Q(t) \propto \lambda^2 \omega_C$.
On the contrary, the QSNR can have a maximum in time and then decay to zero, as shown in panel (a1) for $s=1$, a situation in which the asymptotic probe state has no coherence.
In this case, the parametric plot shows that after the initial trade-off region the QSNR does not increase even if we let the sample absorb more heat.
Otherwise, the QSNR can also tend to a finite value, as shown in panel (b1) for $s=3$.
This behaviour is due to trapped coherences~\cite{Addis2014,Smirne2019}, i.e. the probe state does not become completely dephased asymptotically.
In this case, the parametric plot in panel (b3) shows that after the initial trade-off there may be also regions in which both the absorbed heat and the error decrease simultaneously.
Such behaviors are due to the different time scales over which the two quantities show appreciable variations, as highlighted by the panels (a1) and (b1) on the left of Fig.~\ref{fig:QFIvsHeatPlot1}.
This suggests that the optimal probing strategy calls for a short duration of the interaction, to minimize
both the relative error and the invasiveness on the sample, unless one is in a regime where 
trapped coherence occurs.
In the latter case, in fact, letting the probe and the sample interact longer may lead to an enhanced accuracy, while not increasing further the absorbed heat.
Importantly, the key features of this analysis do not appreciably depend on the cutoff choice, as we show in Appendix~\ref{app:other_spectral_densities_QSNR_vs_heat}.

\emph{Time-optimal schemes.}---
To quantify unambigously the role of time in probing schemes, the total experiment time $\tau$ needs to be treated as a resource.
The probing time $t$ of each experiment can be chosen optimally, corresponding then to a total number of experiments $M = \tau / t$ (assuming the time to prepare the initial state and to perform the measurement is negligible);
this approach is standard in frequency estimation~\cite{Demkowicz-Dobrzanski2015,Smirne2015a,Haase2017}.
Time-optimal quantum thermometry has been studied, but considering a Markovian semigroup evolution~\cite{Correa2015,Sekatski2021}, which may be unfit to capture the short-time dynamics of the probe~\cite{breuer2002theory}.

According to the QCRB, the best accuracy obtainable in a total time $\tau$ is thus 
\begin{equation}
  \label{eq:time_optimalCRB}
    \frac{T^2}{\Delta^2 \tilde{T}} \leq  \tau \max_t \frac{\mathcal{R}_T}{t} \equiv \tau \, \mathfrak{R}_T,
\end{equation}
attainable in the limit $\tau \gg t $ when the experiment is repeated many times.
The optimal QSNR rate $\mathfrak{R}_T$ is the relevant figure of merit for time-optimal schemes; the time $t_\mathrm{opt} = \operatorname{argmax}_t \frac{\mathcal{F}[\rho_T(t)]}{t}$ is the optimal duration of each shot of the experiment and plays an important role.

\begin{figure}[t!]
  \includegraphics{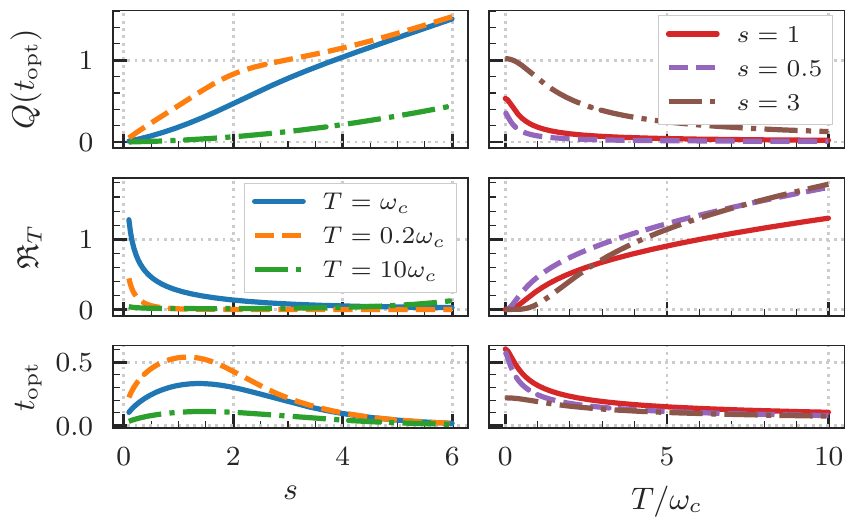}
  \caption{Absorbed heat per shot (top panels), optimal QSNR rate (middle panels) and optimal probing time (bottom panels) as a function of $s$ for $T/\omega_c=0.2,1,10$ (left panels) and as a function of $T$ for $s=1/2,1,3$ (right panels). In both cases $\lambda=1$.
  }
  \label{fig:optQFIvs_s_T}
\end{figure}

We study time-optimal schemes by performing the $t$-optimization in Eq.~\eqref{eq:time_optimalCRB} numerically.
In Fig.~\ref{fig:optQFIvs_s_T} we show the optimal QSNR rate $\mathfrak{R}_T$ and the heat $Q(t_\mathrm{opt})$ absorbed during each shot of the experiment, both as a function of $s$ for three values of $T$ in the left panels, and as a function of $T$ for the three Ohmicity regimes in the right panels.
We also plot the corresponding optimal probing time $t_\mathrm{opt}$.
Notice that $Q(t_\mathrm{opt})$ depends on the temperature implicitly through $t_\mathrm{opt}$.

From the results in Fig.~\ref{fig:optQFIvs_s_T} we see that for decreasing $s$ both  the thermometric accuracy increases and  the absorbed heat decreases. A similar behavior appears for increasing temperature.
Thus, we conclude that time-optimal thermometry is not only beneficial to make the most of the available total time of the experiment, but it is also effective to keep the invasiveness under control.
We also note that small values of $s$ are particularly beneficial in the low-temperature regime, since at the lowest order in $T$ we have $\mathcal{R}_T \propto T^{2(s+1)}$ (see Appendix~\ref{app:dephasingOhm}).

\emph{Role of the coupling strength.}---The very idea behind the use of quantum systems as probes might a priori suggests that a small, albeit indeed non-negligible, coupling strength $\lambda$ should be preferable to reduce the impact of the probe on the sample as much as possible.
However, by taking into account the invasiveness of the probe, we show that this is not necessarily the case.

\begin{figure}[t!]
  \includegraphics{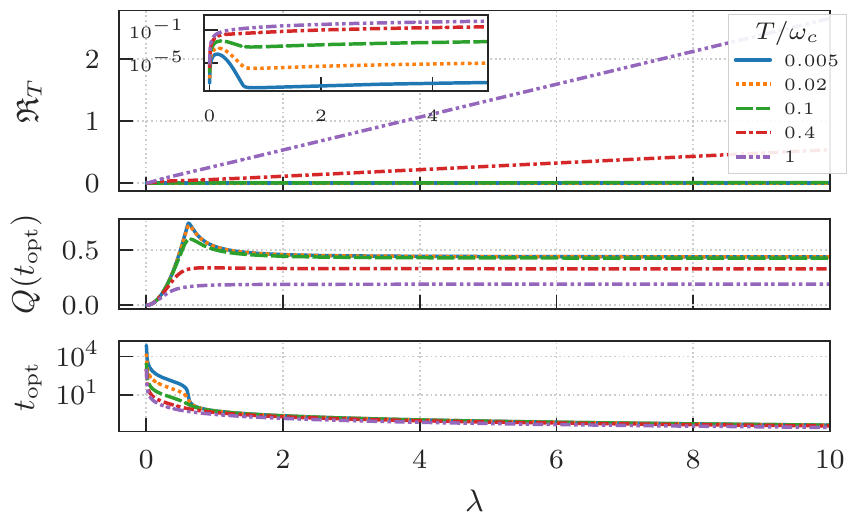}
  \caption{Time-optimal QSNR rate (top panel, the inset is in logarithmic scale), absorbed heat per shot (middle panel),  and optimal time (bottom panel) as a function of $\lambda$ for $s=1$ and several temperature values, shown in the legend. }
  \label{fig:optQFIvs_lambda}
\end{figure}

On the one hand, a stronger coupling increases the amount of heat absorbed by the environment, keeping everything else fixed, since it appears as an overall multiplicative factor in Eq.~\eqref{eq:Qabs_cont}.
On the other hand, it also makes the system lose coherence faster, which means acquiring the information about the temperature faster and thus having a shorter optimal probing time, during which less heat is absorbed.
As shown in Fig.~\ref {fig:optQFIvs_lambda}, the overall behaviour is favorable for large $\lambda$.
While Fig.~\ref{fig:optQFIvs_lambda} shows that the quantities are not monotonic in $\lambda$, as evidenced by the low-temperature curves in the region $0 < \lambda <1$, we see that as $\lambda$ increases the absorbed heat saturates to a constant while the optimal QSNR rate increases linearly with $\lambda$. 
This is caused by the optimal time decreasing sufficently fast as a function of $\lambda$.
While Fig.~\ref{fig:optQFIvs_lambda} shows results for $s=1$, in Appendix~\ref{app:other_spectral_densities_timeopt} we show that the large-$\lambda$ behavior generally holds also for sub-Ohmic and super-Ohmic spectral densities.

Assuming that the optimal time is short, which is true for $\lambda$ large enough, we can expand around $t=0$ both the absorbed heat $Q(t)$ and the QFI.
Quite generally, the first two lowest order terms are quadratic and quartic, a behaviour connected to the so called Zeno regime of open quantum systems~\cite{Chin2012,Macieszczak2015,Smirne2015a}.
Exceptions may arise~\cite{Antoniou2001}, but this holds for the spectral densities we consider.
It is paramount to keep also the fourth order contribution to have a nonmonotonic time-dependence and investigate the behaviour of the optimal probing time~\footnote{This is similar to the way the ``Zeno time'' is defined in Ref.~\cite{Antoniou2001}, i.e. the time when the magnitude of the quartic and quadratic term coincide.}. 
From this optimization we obtain that $t_\mathrm{opt} \propto 1/\lambda$ for large $\lambda$ and thus $Q(t)$ tends to a constant as $\lambda$ increases, while $\mathfrak{R}_T$ grows linearly, see Appendix~\ref{app:short_time} for details.
While the role of coupling strength on the accuracy of quantum thermometry was studied for a few models~\cite{Correa2017,Mitchison2020} showing different behaviours, no considerations about time-optimality were previously made.

Furthermore, we can apply these results to a spin-$j$ probe, coupled to the sample via the operator $A_S = 2 \lambda J_z$ and initialized in a spin cat state $(\ket{j,j}+\ket{j,-j})/\sqrt{2} $.
The dynamics is then equivalent to a spin-$\frac{1}{2}$ probe under the scaling $ \lambda \mapsto 2j \lambda$, but in this physical setting $j$ may be increased instead of $\lambda$.
However, for higher spin the choice of such an initial probe state is not obvious.
In Appendix~\ref{app:GHZ} we show that, in some regimes, spin cat states are optimal and match the ultimate metrological bounds for quantum channels~\cite{Fujiwara2008,Demkowicz-Dobrzanski2012}.
In particular, spin cat probes become optimal when $\lambda$ is large, thus we suspect they may be optimal also when $j$ is large (for a fixed $\lambda$), beyond what we can reach with our numerics.

\emph{Discussion.}---In this Letter we have analysed a dephasing thermometry protocol, taking into consideration both the information encoded in the probe and the transformation of the sample due to the interaction with the probe, i.e., the invasiveness of the probing scheme.
Invasiveness has been studied in optical quantum metrology by imposing constraints on the amount of light absorbed by the sample~\cite{Perarnau-Llobet2020a}, e.g. fragile biological ones~\cite{Taylor2016}.
In a similar fashion, we have quantified the invasiveness in quantum thermometry with the amount of heat absorbed by the sample.

We have shown that in time-optimal schemes the relative error in temperature estimation and the invasiveness due to the  probe-sample interaction exhibit the same behavior as function of the model parameters, thus indicating that
strategies minimizing the former also limit the latter.
On the other hand, in the strong-coupling regime the relative error becomes inversely proportional to the coupling, while the absorbed heat per shot reaches a constant value, meaning that a strong interaction between the quantum probe and the sample can lead to the most effective probing strategy.

Going beyond pure dephasing, it will be interesting to study the invasiveness of thermometry schemes in which the probe's energy can change, e.g. using quantum thermal machines as thermometers~\cite{Hofer2017d,Henao2021a}.
Moreover, invasiveness could also be characterized taking into account energy fluctuations or by considering the post-measurement state~\cite{Seveso2018} and extended to thermometry with sequential measurements on the probe~\cite{Burgarth2015a,DePasquale2017}.

As a final comment, we stress that energetic efficiency will increasingly become a relevant issue for all quantum technologies~\cite{Auffeves2022}.
The energetic cost of phase and frequency estimation has been explored~\cite{Lipka-Bartosik2018,Liuzzo-Scorpo2018}, but the field is arguably in early stages.
Our approach is also a first step in this direction for quantum thermometry, since the heat absorbed by the environment coincides with the external work for coupling and decoupling the probe for the pure dephasing model we have considered.

\emph{Acknowledgments.}---We thank C. Benedetti, A. Candeloro and A. Saltini for fruitful discussions.
The authors acknowledge financial support from MUR under the ``PON Ricerca e Innovazione 2014-2020'' project EEQU.

\bibliography{thermo_bib}

\clearpage

\appendix
\setcounter{figure}{0}
\setcounter{page}{1}
\setcounter{equation}{0}
\renewcommand{\thefigure}{S\arabic{figure}}
\renewcommand{\theequation}{S\arabic{equation}}
\renewcommand{\appendixname}{}

\widetext
\section{Pure dephasing of a qudit interacting with a bath of harmonic oscillators}
\label{app:puredephasing}

\subsection{General solution}

\subsubsection{Global system-environment unitary evolution operator}
We assume the following system Hamiltonian, written in the energy eigenbasis:
\begin{equation}
    H_S = \sum_j \epsilon_j | j \rangle \langle j |.
\end{equation}
We assume a standard bath of harmonic oscillators with the free Hamiltonian
\begin{equation}
  H_E = \sum_k \omega_k b^\dag_k b_k.
\end{equation}
We consider a pure dephasing evolution of the system, in which the interaction term commutes with $H_S$ and is diagonal in the energy eigenbasis:
\begin{equation}
  \label{eq:HSB_schropic}
  H_{I}^{\mathrm{Sch}} = \left( \sum_j g_j | j \rangle \langle j| \right) \otimes \left( \sum_k f_k b^\dag_k + f_k^* b_k \right).
\end{equation}
Going into the interaction picture we obtain (the system term of the interaction Hamiltonian is unchanged because of the commutativity assumption)
\begin{equation}
  \label{eq:int_pic_Hamiltonian}
  H_{I}(t) = \sum_j g_j | j \rangle \langle j | \otimes \left( \sum_k  f_k e^{ \I \omega_k t } b^\dag_k  + f_k^* e^{- \I \omega_k t } b_k   \right).
\end{equation}
The commutator at different times is proportional to the identity on the oscillators:
\begin{equation}
  [H_{I}(t), H_{I}(t')] = -2 \I \sum_j g_j^2 \left[ \sum_k |f_k|^2 \sin \omega_k (t-t') \right]   | j \rangle \langle j| \otimes \id,
\end{equation}
since it commutes with the interaction Hamiltonian at all time the evolution operator in the interaction picture can be obtained exactly since the Magnus expansion terminates at second order, obtaining:
\begin{equation}
  \label{eq:U_IgeneralDephasing}
  \begin{split}
  U_I(t) &= \exp \left[ - \frac{1}{2} \int_0^t ds \int_0^s ds' [H_I(s),H_I(s')] \right] \exp \left[ -\I \int_0^t ds H_I(s) \right] \\
  & = \sum_j \exp \Biggl[ \I 
  \underbrace{\sum_k |f_k|^2 g_j^2 \frac{\omega_k t - \sin (\omega_k t)}{\omega_k^2}}_{\textstyle \equiv \phi_j^{\mathrm{int}}(t)} \,\Biggr] |j \rangle \langle j| \otimes \exp \left[ \sum_k \alpha_k^j(t) b_k^\dag - \alpha_k^{j*}(t) b_k \right], \\ 
  \end{split}
\end{equation}
where the action on the oscillators is a product of single-mode displacements
\begin{equation}
  \exp \left[ \sum_k \alpha_k^j(t) b_k^\dag - \alpha_k^{j*}(t) b_k \right] = \prod_k D(\alpha_k^j(t)) \quad \text{with} \quad \alpha_k^{j}(t)= -\I f_k g_j \int_0^t \! ds \, e^{\I \omega_k s} = f_k g_j \frac{1 - e^{\I \omega_k t }}{\omega_k},
  \label{eq:alphat}
\end{equation}
where $ D(\alpha) = \exp \left[ \alpha b^\dag - \alpha^{*} b \right]$.
Thus, the overall evolution is a displacement of the oscillators conditioned on the state of the system.

\subsubsection{Reduced state of the system}

We write a generic initial state as $\rho_{SE}(0) = \sum_{ij} \rho_{S,ij}(0) | i \rangle \langle j |  \otimes \rho_E^{ij}$, where $ \Tr_E \rho_E^{ij} = 1$ are normalized operators, but not necessarily states when $i\neq j$.
When $i=j$ these are the conditional states obtained with probability $\rho_{S,ii}(0)$ by measuring the system in the basis $\ket{j}$.
This means that the initial reduced state of the system is $\rho_{S}(0) = \Tr_E \rho_{SE}(0) = \sum_{ij} \rho_{S,ij}(0) | i \rangle \langle j | $.
We can thus write the evolved state as 
\begin{equation}
  \rho_{S}(t) =  \sum_{i} \rho_{S,ii}(0) | i \rangle \langle i | + \sum_{i \neq j} \rho_{S,ij}(0) \Tr_{E}\left[ \prod_k D(\alpha_k^i(t)) \rho_E^{ij} \prod_{k'} D(-\alpha_{k'}^j(t)) \right] | i \rangle \langle j |,\label{eq:solg}
\end{equation}
showing explicitly that the populations are constants of motion since $\Tr_{E}\left[ \prod_k D(\alpha_k^i(t)) \rho_E^{ii} \prod_{k'} D(- \alpha_{k'}^i(t)) \right] = 1 $ is the trace of a normalized state.
On the contrary the off-diagonal elements are changed by a factor $\Tr_{E}\left[ \prod_k D(\alpha_k^i(t)) \rho_E^{ij} \prod_{k'} D(-\alpha_{k'}^j(t)) \right]$.
Note that (\ref{eq:solg}) includes the possibility of having initial system-environment correlations---see also \cite{Morozov2012} for a study of pure dephasing with correlated initial states.

From now on we assume an initial product state for the system and environment $\rho(0) \otimes \nu_T$, where the state of the environment $\nu_T = \bigotimes_k \nu_k$ is a thermal state, factorized into thermals states of each mode since the oscillators are not interacting.
The evolved global state is
\begin{equation}
  \rho_{SE}(t)  = U_I(t) \left( \rho_S(0) \otimes \nu_T \right) U_I(t)^\dag = \sum_{ij} \rho_{S,ij}(0) e^{\I(\phi_i^{\mathrm{int}}(t) - \phi_j^{\mathrm{int}}(t))} |i \rangle \langle j | \otimes \prod_k D( \alpha_k^{i}(t)) \nu_T D( -\alpha_k^{j}(t)).
\end{equation}
To evaluate the trace on the environment we use the Baker-Campbell-Haussdorff formula to obtain
\begin{equation}
\label{eq:TrEk_Deph}
\Tr_{k} \left[  D( \alpha_k^{i}(t)) \nu_{k} D( -\alpha_k^{j}(t)) \right] = \Tr_{k} \left[  D( \alpha_k^{i}(t) -\alpha_k^{j}(t) ) \nu_k \right] \exp \left[ \frac{1}{2} \left( -\alpha_k^{j}(t)\alpha_k^{* i}(t) + \alpha_k^{*j}(t)\alpha_k^{i}(t) \right) \right],
\end{equation}
where $\Tr_{k} \left[  D( \alpha ) \nu_k \right]= \frac{1}{2}|\alpha|^2 \coth \left( \frac{\omega_k}{2 T}\right) $ is the characteristic function of a thermal state.

Overall, keeping track of all the phase factors, the reduced system state in Schrödinger picture is thus
\begin{equation}
  \rho_S(t) = \Tr_E \left[ \rho_{SE} (t)  \right] = \sum_{ij} \rho_{S,ij}(0) e^{ - \left( \Delta_{ij}(t) + \I \varphi_{ij}(t) \right) } |i \rangle \langle j |, 
\end{equation}
where the real-valued dephasing function affecting the off-diagonal elements is
\begin{equation}
  \Delta_{ij}(t) = - \sum_k \ln \Tr_{k} \left[  D\left( \alpha_k^i(t) - \alpha_k^j(t)  \right)  \nu_T \right] = \sum_k \frac{1}{2} \left| \alpha_k^i(t) - \alpha_k^j(t)  \right|^2 \coth \left( \frac{\omega_k}{2 T} \right),
\end{equation}
while the phase factor includes the usual difference of unitary phases $\phi^\mathrm{Sch}_j(t) = \phi^{\mathrm{int}}_j(t) - t \epsilon_j $, containing both the effect of the system Hamiltonian and the phases appearing in the interaction-picture unitary~\eqref{eq:U_IgeneralDephasing}, so that the phase factor mentioned in the main text reads $\varphi_{ij}(t) = \phi^\mathrm{Sch}_j(t) - \phi^\mathrm{Sch}_i(t)$.
Besides being obviously irrelevant for energetic considerations on the system, the phase factors do not depend on the environment initial state and thus on the temperature, so they are also irrelevant for the QFI.
However, they may be useful to learn properties of the environment spectral density, see e.g.~\cite{Wang2021a}.
Taking the continuous limit $ \sum_k |f_k|^2 \mapsto \int_0^\infty d \omega J(\omega) $ and using the definition \eqref{eq:alphat} we obtain Eq.~\eqref{eq:Gamma_ij_cont} in the main text.

\subsubsection{Heat absorbed by the environment}
The reduced state of the environment is the mixture
\begin{equation}
 \rho_B(t) = \Tr_S \left[ \rho_{SB} (t)  \right] =  \sum_j  \rho_{S,jj}(0) \prod_k D( \alpha_k^{j}(t)) \nu_T D( -\alpha_k^{j}(t)).
\end{equation}
We can evaluate the energy of the bath (we can use the interaction-picture operator, since the transformation back to the Schrödinger picture commutes with the free Hamiltonian)
\begin{equation}
  \begin{split}
  \langle H_B(t) \rangle &= \Tr_B \left[ H_B \rho_B(t) \right] =  \sum_j  \rho_{S,jj}(0)  \sum_k \omega_k \Tr_{k}\left[  b_k^\dag b_k D(\alpha_k^j(t)) \nu_k D(-\alpha_k^j(t)) \right] \\
  &= \sum_j \rho_{S,jj}(0) \sum_k \omega_k \left( \frac{1}{e^{\omega_k/T}-1} + \left| \alpha_k^j(t) \right|^2  \right).
  \end{split}
\end{equation}
The absorbed heat is thus
\begin{equation}
 Q(t) = \langle H_B(t) \rangle  - \langle H_B(0) \rangle = \sum_j \rho_{S,jj}(0) \sum_k \omega_k \left| \alpha_k^j(t) \right|^2 ,
\end{equation}
where we see that the thermal contribution remains the same and the absorbed energy only depends on the displacement due to the interaction with the qudit.
Taking the continuous limit and using the definition \eqref{eq:alphat} we obtain Eq.~\eqref{eq:Qabs_cont} in the main text.

\subsection{Ohmic-like spectral density}
\label{app:dephasingOhm}

\subsubsection{Dephasing function}
For a spectral density with exponential cutoff, we obtain the dephasing function (this expression is slightly different from Eq.~(17) in Ref.~\cite{Razavian2018}, due to some typos therein)
\begin{align}
  \label{eq:GammaTOhmicAnalytical}
  \Delta_{0}(t) &= \Gamma (s-1) \left\{1-\left(\tilde{t}^2+1\right)^{\frac{1}{2}-\frac{s}{2}} \cos \left[(s-1) \arctan(\tilde{t})\right]\right\} \\
  \Delta_{T}(t) &= \Delta_{0}(t)+\frac{(s-1) s \tilde{T}^{s-1} \Gamma (s-1)^2}{\Gamma (s+1)} \biggl[ 2 \zeta (s-1,\tilde{T}+1) -\zeta (s-1, \I \tilde{t} \tilde{T} + \tilde{T} + 1) - \zeta (s-1,-\I \tilde{t} \tilde{T}+\tilde{T}+1) \biggr],
\end{align}
where $\zeta(s,a)=\sum_{n=0}^\infty \frac{1}{(n+a)^s}$ is the generalized (Hurwitz) zeta function and we have introduced the adimensional $\tilde{t}=t\omega_c$ and $\tilde{T}=T/\omega_c$ for compactness.
Also note that in Ref.~\cite{Razavian2018} the spectral density $J(\omega)$ is defined with an addictional factor 4 and the results shown there are obtained by fixing $\lambda=1/2$ in this work.
For the other cutoff functions we were able to evaluate the dephasing function analytically only for $T=0$, which corresponds to the calculation of the absorbed heat in the next section.

From the dephasing function we can evaluate the QFI as
\begin{equation}
  \mathcal{F}[\rho_T(t)] = \frac{ 16 \lambda^4 \, \left[ \partial_T \! \Delta_T(t)\right]^2 }{\exp\left[ 8 \lambda^2 \Delta_T(t) \right] -1 }.
\end{equation}
The full expression is involved and not particularly instructive, however the leading order term for $T \to 0$ reads
\begin{equation}
  \mathcal{F}[\rho_T(t)]  = \frac{16 \lambda ^4 s^2 (s+1)^2 t^4 \omega_c^2 \zeta (s+1)^2 \Gamma (s)^2 \left(\frac{T}{\omega_c}\right)^{2 s}}{\exp \left(8 \lambda ^2 \Gamma (s-1) \left(1-\left(t^2 \omega_c^2+1\right)^{\frac{1}{2}-\frac{s}{2}} \cos \left((s-1) \tan ^{-1}(t \omega_c)\right)\right)\right)-1},
\end{equation}
showing that it goes to zero more slowly for $ T / \omega_c \ll 1$ as $s \to 0$.
However, the QFI always goes to zero as $ T \to 0$ for $s > 0$ and thus the absolute error diverges.
We note that for other probe-sample interactions one can actually have a vanishing absolute error in the limit $T \to 0$, while the relative error must diverge~\cite{Jorgensen2020}.

\subsubsection{Absorbed heat}
\begin{figure}
  \includegraphics[width=.32\textwidth]{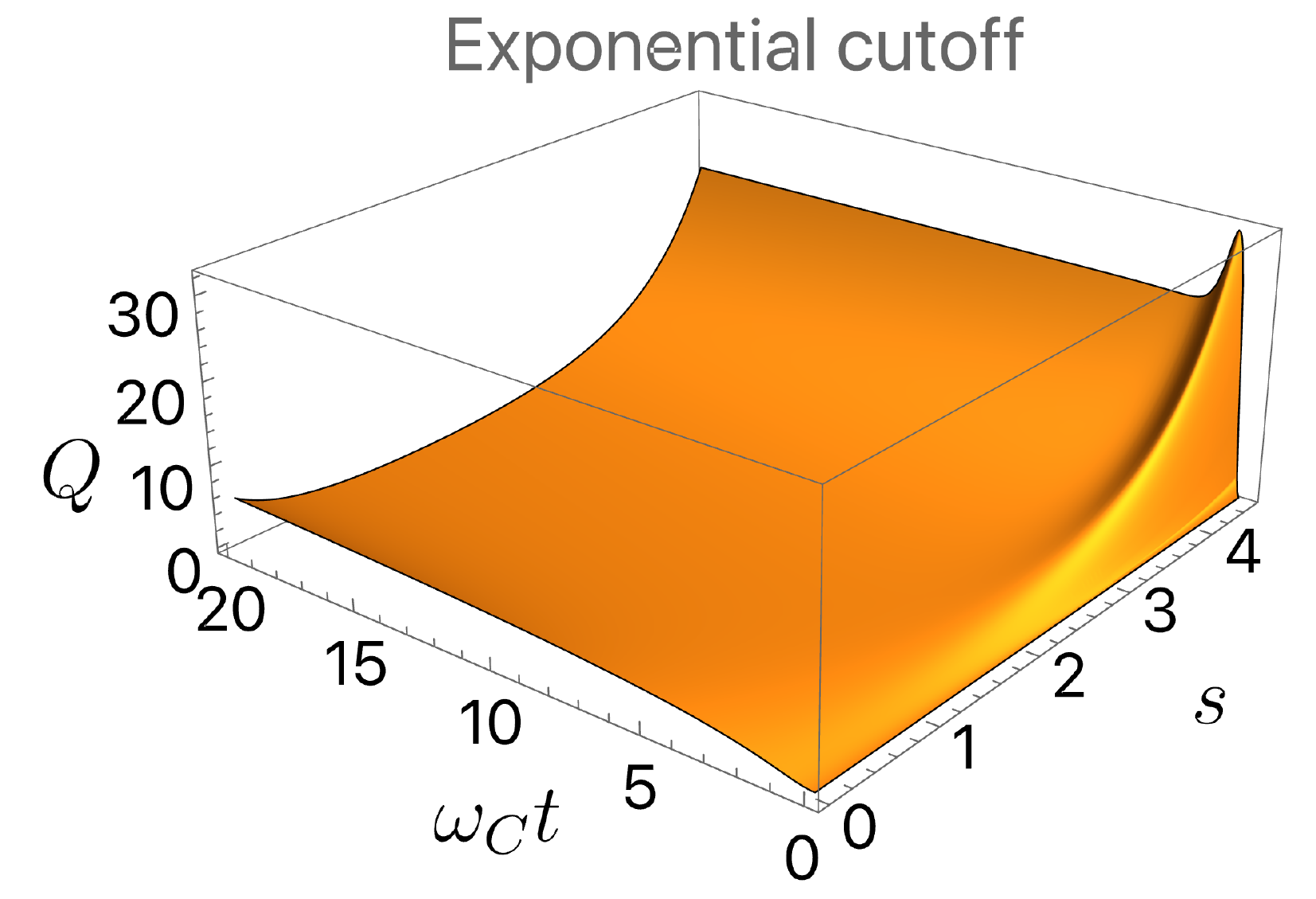}
  \includegraphics[width=.32\textwidth]{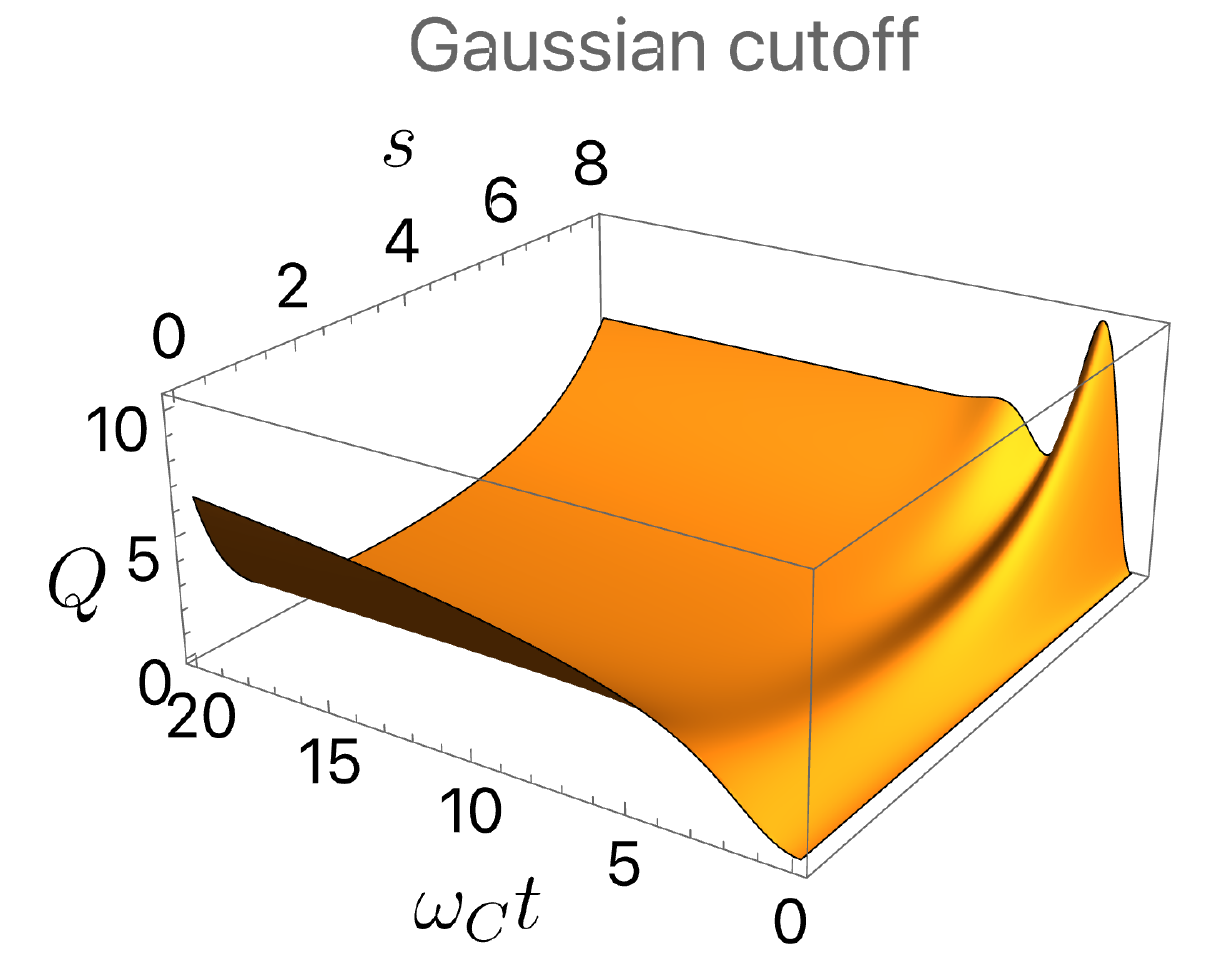}
  \includegraphics[width=.32\textwidth]{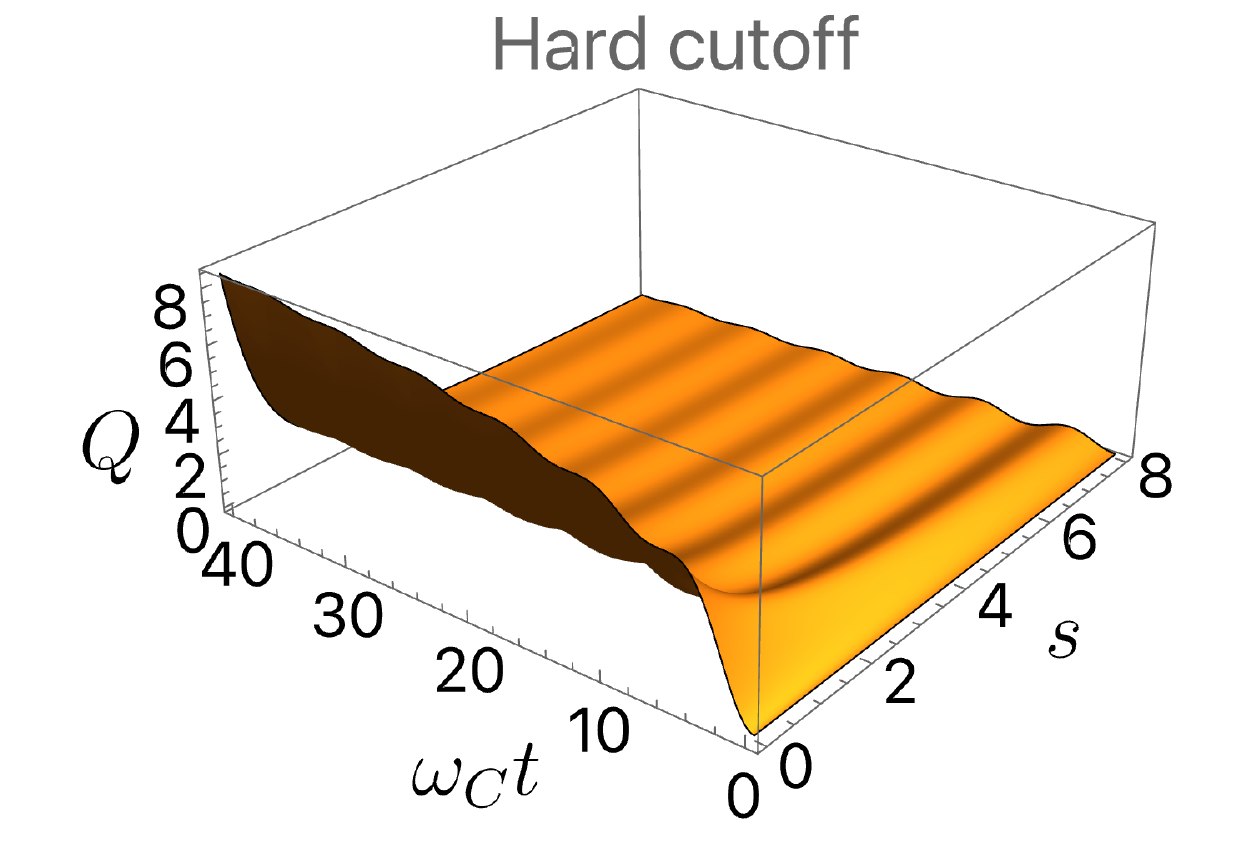}
  \caption{Absorbed heat for Ohmic-like spectral densities, as a function of evolution time and Ohmicity parameter $s$, for three choices of cutoff functions.}
  \label{fig:absQ_3D}
\end{figure}

For the considered model the absorbed heat depends non-trivially only on the evolution time and on the Ohmicity parameter, since $\lambda^2$ is only a multiplicative factor and there is no temperature dependence, as can be seen from Eq.~\eqref{eq:Qabs_cont}. This function of two parameters is shown in Fig.~\ref{fig:absQ_3D} for different cutoff functions.
For the exponential cutoff the explicit expression is reported in Eq.~\eqref{eq:QabsExpCutoff} of the main text, and we see that it tends to increase for large $s$ at all times, while showing a peak for short times only in the super-Ohmic region.
We also see that the asymptotic value is not monotonic in $s$, since
\begin{equation}
  \lim_{t\to\infty} \mathbbm{Q}^\mathrm{exp}(t) = 2 \lambda ^2 \omega_c \Gamma (s),
\end{equation}
and it is minimal at the minimum of the Gamma function $s_0 \approx 1.4616$.

Also for the other cutoff functions the absorbed heat can be evaluated analytically.
For a Gaussian cutoff $C(\omega,\omega_c)=e^{-\omega^2 / \omega_c^2}$ we obtain
\begin{equation}
  \mathbbm{Q}^\mathrm{Gauss}(t) = \lambda ^2 \omega _c \Gamma \left(\frac{s}{2}\right) \left(1-\, _1F_1\left( \left[\frac{s}{2}\right]; \left[\frac{1}{2}\right];-\frac{1}{4} t^2 \omega _c^2\right)\right),
\end{equation}
with the asymptotic value
\begin{equation}
  \lim_{t\to\infty} \mathbbm{Q}^\mathrm{Gauss}(t) = \lambda ^2 \omega_c \Gamma \left(\frac{s}{2}\right)
\end{equation}
For the hard cutoff $C=\Theta( \omega_c - \omega )$, where $\Theta(s)$ is the Heaviside step function, we obtain 
\begin{equation}
  \mathbbm{Q}^\mathrm{hard}(t) = \frac{2 \lambda ^2 \omega _c \left[1 - \, _1F_2\left( \left[\frac{s}{2} \right]; \left[\frac{1}{2},\frac{s}{2}+1\right];-\frac{1}{4} t^2 \omega _c^2\right) \right]}{s},
\end{equation}
where $\,_p F_q( \vec{a};\vec{b};z) = \sum_{k=0}^{\infty } \frac{  (a_1)_k \ldots    (a_p)_k}{(b_1)_k \ldots (b_q)_k} \frac{z^k}{k!}$ is the generalized hypergeometric function and $(a)_k=\prod_{j=1}^k(a+j-1)$ is the Pochhammer symbol.
The corresponding asymptotic value is
\begin{eqnarray}
  \lim_{t\to\infty} \mathbbm{Q}^\mathrm{hard}(t) = \frac{2\lambda ^2 \omega_c}{s} .
\end{eqnarray}
The non-trivial dependence on the parameters $s$ and $t$ for these other two cutoff functions is shown in Fig.~\ref{fig:absQ_3D}.
There are several qualitative differences between the cutoff functions, the most important being that for a hard cutoff we see oscillations and the absorbed heat tends to zero for large $s$.

\section{Additional results for other spectral densities}
\label{app:other_spectral_densities}
In this appendix we present additional plots for different spectral densities than those considered in the main text.
We explore Gaussian and hard cutoff functions and different Ohmicity parameters.
Overall, we observe that the key qualitative observations presented in the main text remain valid, while some finer details depend on these features of the spectral density.

\subsection{Thermometric performance versus heat as a function of time}
\label{app:other_spectral_densities_QSNR_vs_heat}
\begin{figure*}[ht!]
  \centering
  \includegraphics{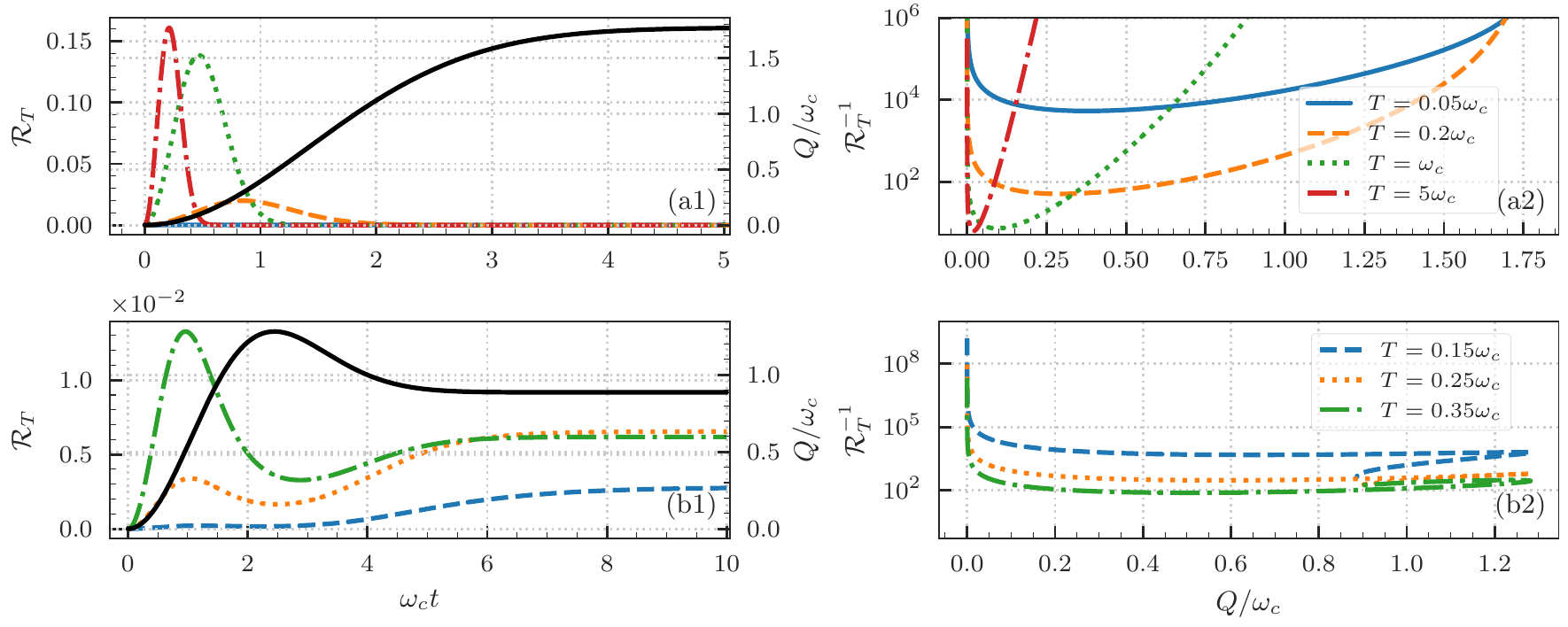}
  \caption{Plots analogous to Fig.~\ref{fig:QFIvsHeatPlot1} in the main text, but for Ohmic-like spectral densities with a Gaussian cutoff.
  Top [panels (a)]: Ohmic spectral density $s=1$; 
  bottom [panels (b)]: super-Ohmic spectral density $s=3$; both for coupling $\lambda = 1$.
  Panels (a1) and (b1): absorbed heat (solid black line, units on the right) and quantum SNR (coloured lines for various temperatures, units on the left) as a function of time. 
  Panels (a2) and (b2): parametric plot of relative error for temperature estimation versus absorbed heat.
  Each line represents a different temperature, as shown in the legend.
  }  
\label{fig:QFIvsHeatPlotGauss}
\end{figure*}

\begin{figure*}[ht!]
  \centering
  \includegraphics{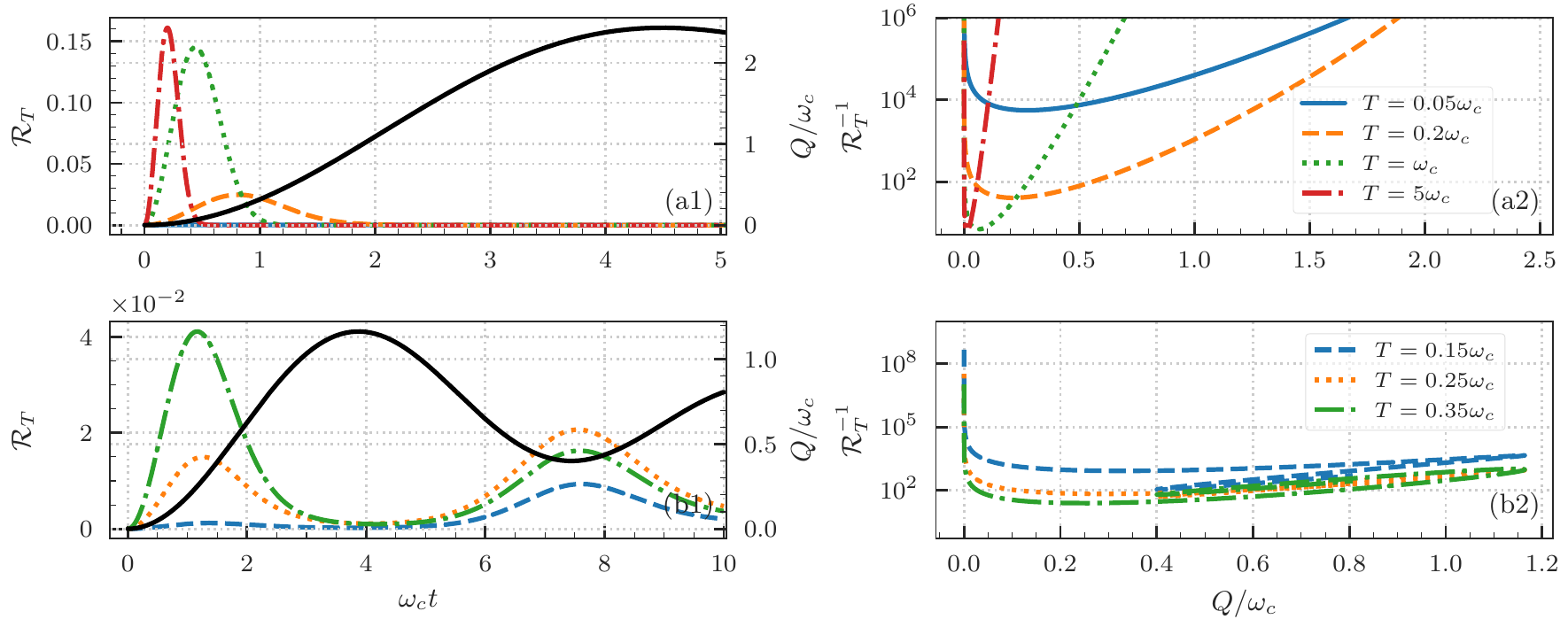}
  \caption{Plots analogous to Fig.~\ref{fig:QFIvsHeatPlot1} in the main text, but for Ohmic-like spectral densities with a hard (step function) cutoff.
  Top [panels (a)]: Ohmic spectral density $s=1$; 
  bottom [panels (b)]: super-Ohmic spectral density $s=3$; both for coupling $\lambda = 1$.
  Panels (a1) and (b1): absorbed heat (solid black line, units on the right) and quantum SNR (coloured lines for various temperatures, units on the left) as a function of time. 
  Panels (a2) and (b2): parametric plot of relative error for temperature estimation versus absorbed heat.
  Each line represents a different temperature, as shown in the legend.
  }  
\label{fig:QFIvsHeatPlotHard}
\end{figure*}

In Fig.~\ref{fig:QFIvsHeatPlotGauss} and~\ref{fig:QFIvsHeatPlotHard} we reproduce the plot in Fig.~\ref{fig:QFIvsHeatPlot1} in the main text, but for a Gaussian and a hard cutoff, respectively.
The main qualitative features are the same.
There is an initial tradeoff between absorbed heat and QSNR at short times, since both quantities start from zero.
The heat does not decrease back to zero asymptotically, while the QNSR does for $s=1$, but not for $s=3$ when trapped coherences are present in the probe.
The main qualitative difference is that for a hard cutoff an oscillatory behaviour for sufficiently long times can be observed, which is not present for Gaussian and exponential cutoffs.

\subsection{Time-optimal thermometry}
\label{app:other_spectral_densities_timeopt}

\begin{figure}[ht!]
  \includegraphics{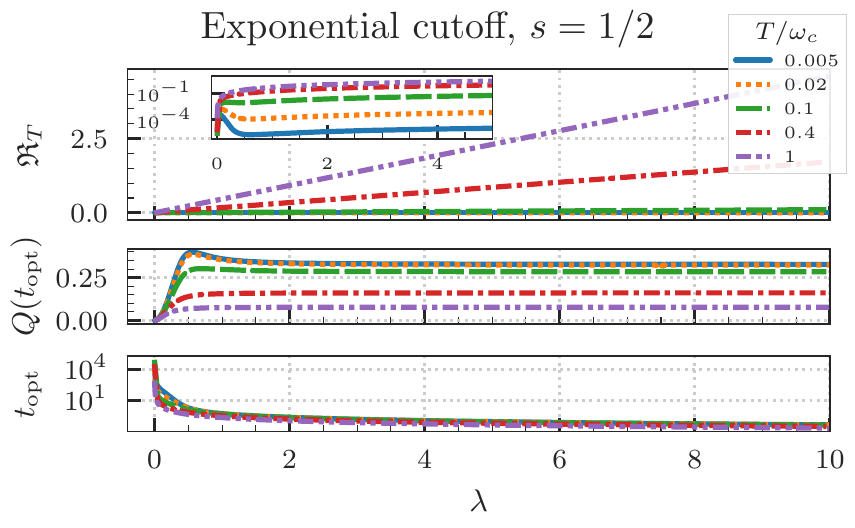}
  \includegraphics{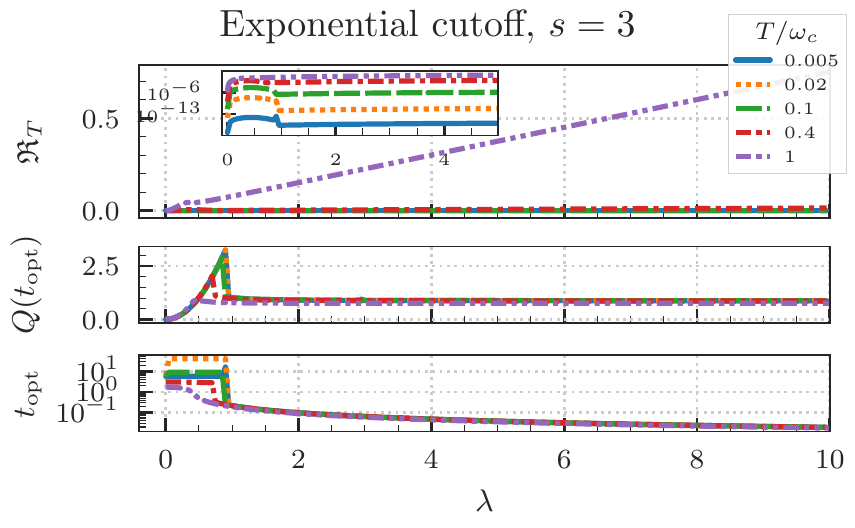}
  \caption{Plots analogous to Fig.~\ref{fig:optQFIvs_lambda} in the main text, but for $s=1/2$ (left) and $s=3$.
  Time-optimal QSNR rate (top panels, the insets are in logarithmic scale), absorbed heat per shot (middle panels),  and optimal time (bottom panels), all plotted as a function of $\lambda$ for several temperature values, shown in the legend. }
  \label{fig:optQFIvs_lambda_subsuperOhmic}
\end{figure}

In Fig.~\ref{fig:optQFIvs_lambda_subsuperOhmic} we reproduce the plots of Fig.~\ref{fig:optQFIvs_lambda} in the main text, but for a sub-Ohmic spectral density $s=1/2$ and for a super-Ohmic spectral density $s=3$.
We see that the large-$\lambda$ behaviour highlighted in the main text, i.e. $Q(t_\mathrm{opt})$ saturating to a finite value and $\mathfrak{R}_T$ growing linearly, remains valid.
However, we see that the behaviour for smaller values of $\lambda$ are rather different, with the super-Ohmic case showing more peculiar features.

\begin{figure}[ht!]
  \includegraphics{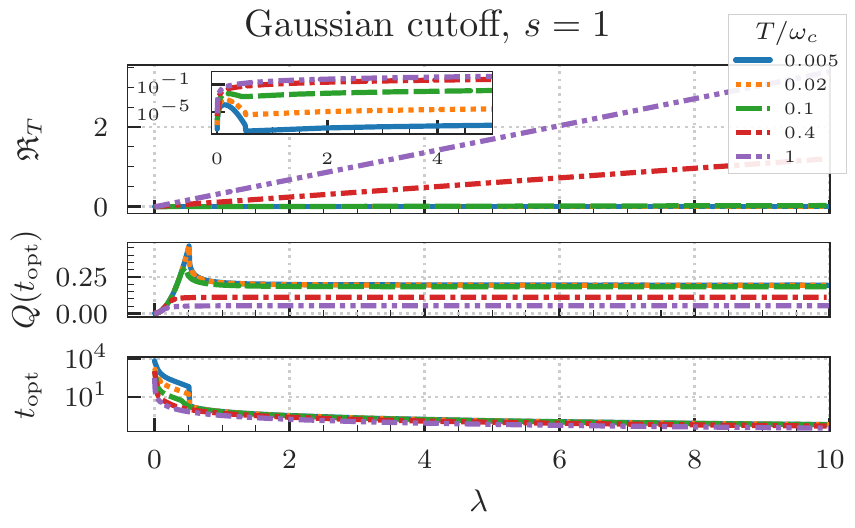}
  \includegraphics{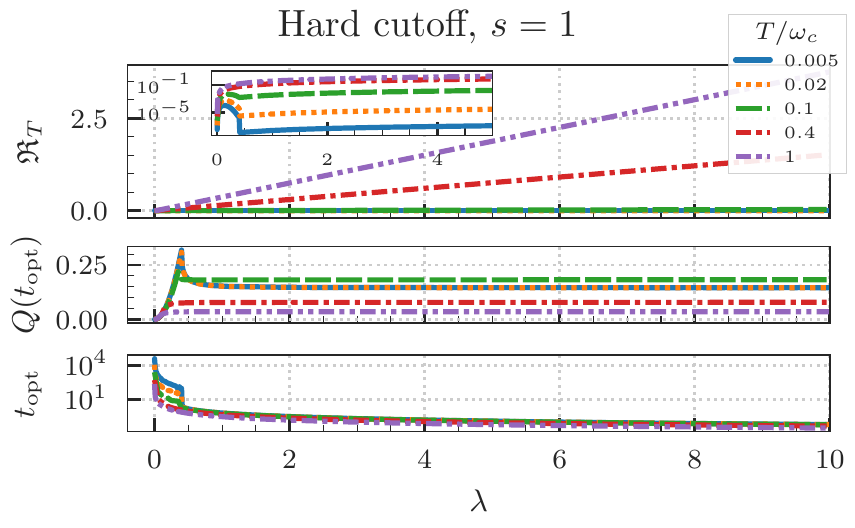}
  \caption{Plots analogous to Fig.~\ref{fig:optQFIvs_lambda} in the main text, but for a Gaussian (left) and hard (right) cutoff, both for $s=1$ as in Fig.~\ref{fig:optQFIvs_lambda}.
  Time-optimal QSNR rate (top panels, the inset is in logarithmic scale), absorbed heat per shot (middle panels),  and optimal time (bottom panels) as a function of $\lambda$ for several temperature values, shown in the legend. }
  \label{fig:optQFIvs_lambda_Gauss_hard}
\end{figure}

In Fig.~\ref{fig:optQFIvs_lambda_Gauss_hard} we reproduce again the plot in Fig.~\ref{fig:optQFIvs_lambda} in the main text, keeping the Ohmicity parameter $s=1$, but for Gaussian and hard cutoffs

\section{Numerical evidence for the optimality of spin cat states}
\label{app:GHZ}

Before considering all the details of the temperature estimation problem in the main text, we need to consider the underlying problem of estimating a constant dephasing factor.

\subsection{Evaluation of the optimal QFI for dephasing estimation}
We focus on the estimation of a \emph{constant} dephasing factor $\Delta$ characterizing a dephasing channel $\mathcal{E}_\Delta$ that acts as follows on a finite-dimensional system
\begin{equation}
  \label{eq:dephasing_est_Appendix}
  \mathcal{E}_\Delta[ \rho ] = \sum_{i,j} \rho_{ij} e^{- \Delta (i-j)^2} | i \rangle \langle j | \equiv  \mathbf{E}_\Delta \circ \rho  \qquad  \mathbf{E}_\Delta = \sum_{i,j} e^{- \Delta (i-j)^2} | i \rangle \langle j| 
\end{equation}
where $\circ$ denotes the elementwise (Hadamard) product between two matrices.
This channel encodes the ``operatorial'' part of the thermometry problem considered in the main text.
The fact that $\Delta$ is actually a time-dependent function of the temperature appears in the QFI only as a multiplicative factor, which mainly plays a role in the optimization over the probing time.
The dephasing matrix $\mathbf{E}_\Delta$ is essentially the Choi matrix of the channel, after removing redundant columns and rows of zeros.

To evaluate the optimal QFI we use the method introduced in Ref.~\cite{Fujiwara2008} based on the optimization over equivalent Kraus representations $\{ K_j \}_{j=1,\dots,r}$ of the dynamical map $\mathcal{E}_\Delta[ \cdot ] =  \sum_{k=1}^r K_k \cdot K_k^\dag$ that encodes the parameter.
Explicitly, it can be evaluated as the minimization over a hermitian matrix of size $r \times r$ of a quadratic function of $h$ involving the Kraus operators and their derivatives: 
\begin{equation}
  \label{eq:channelQFI}
  \mathfrak{F}(\Delta) = \max_{\ket{\psi}_{SA}} \mathcal{F}\left[ \mathcal{E}_\Delta \otimes \id_A [ \ket{\psi}_{SA} ] \right] = 4 \min_{h=h^\dag} \left\Vert \sum_k \left( \dot{K}_k - \I \sum_j h_{kj} K_j \right)^\dag \left( \dot{K}_k - \I \sum_{j'} h_{kj'} K_{j'} \right) \right\Vert.
\end{equation}
Here we see that the quantity evaluated by this method is not only an optimization over initial states of the system, but also includes the possibility of using a noiseless ancillary system of arbitrary dimension and initial entangled states.
If the noiseless ancillary system is not available the quantity in Eq.~\eqref{eq:channelQFI} is generally just an upper bound.
Crucially, this minimization can be evaluated numerically by solving a semidefinite program~\cite{Demkowicz-Dobrzanski2012}.

Given a spectral decomposition of the (real, positive semidefinite) dephasing matrix  $\mathbf{E}_\Delta = \sum_{j=1}^{d} \kappa_j \mathbf{k}_j \mathbf{k}_j^T$, one can write a Kraus representation made of diagonal operators
\begin{equation}
  \label{eq:dephasing_Kraus}
   K_j = \sqrt{\kappa_j}\mathrm{diag}(\mathbf{k}_j).
\end{equation}
Since the derivative of the dephasing matrix $\dot{\mathbf{E}}\equiv \partial_\Delta \mathbf{E}_\Delta$ is known:
\begin{equation}
  \partial_\Delta \mathcal{E}_\Delta[ \rho ] = \sum_{i\neq j} \rho_{ij} (i-j)^2 e^{- \Delta (i-j)^2} | i \rangle \langle j | \equiv  \dot{\mathbf{E}}_\Delta \circ \rho  \qquad   \dot{\mathbf{E}} = - \sum_{i \neq j} (i-j)^2 e^{- \Delta (i-j)^2} | i \rangle \langle j|,
\end{equation}
to compute the derivatives the Kraus operators we need to evaluate the derivatives of the eigevnalues $\{ \kappa_j\}$ and eigenvectors $\{ \mathbf{k}_j \}$ through first-order perturbation theory:
\begin{equation}
  \label{eq:dephasing_Kraus_deriv}
  \dot{\kappa_j} = \mathbf{k}_j^T \; \dot{\mathbf{E}} \; \mathbf{k}_j, \quad
  \dot{\mathbf{k}}_j = \sum_{i\neq j} \frac{ \mathbf{k}_j^T \; \dot{\mathbf{E}} \; \mathbf{k}_i }{\kappa_i - \kappa_j}, \; \qquad \dot{K}_j = \frac{ \dot{\kappa_j}}{2 \sqrt{\kappa_j}} \mathrm{diag}(\mathbf{k}_j) + \sqrt{\kappa_j}\mathrm{diag}(\dot{\mathbf{k}}_j).
\end{equation}

Summing up, we can evaluate the optimal QFI numerically by first diagonalizing the dephasing matrix in Eq.~\eqref{eq:dephasing_est_Appendix}, from which the Kraus operators and their derivatives can be evaluated through Eqs.~\eqref{eq:dephasing_Kraus} and~\eqref{eq:dephasing_Kraus_deriv}, in turn these two sets of operators are fed to a semidefinite program that solves the minimization in Eq.~\eqref{eq:channelQFI}.

While we have used this method for quantum thermometry, it could be applied to other estimation problems.
For example, in Ref.~\cite{Rossi2015} the similar problem of estimating a parameter appearing in the dephasing function of a many-qubit state was studied by numerically sampling random probe states, showing that in some regimes GHZ states (completely analogous to spin cat states, but considering multiqubit systems instead of a single spin-$j$ system) are optimal, similarly to what we show next.

\subsection{Comparison between time-optimal schemes with spin cat and optimal probe states}

By employing the optimal QFI presented in the previous section to perform the time-optimization we obtain the ultimate performance achievable with spin-$j$ probe states.
Numerically we evaluate the following quantity:
\begin{equation}
  \label{eq:numerical_topt}
  \max_t \frac{1}{t} \left[ 4 \lambda^2 \partial_T \Delta_T(t) \right]^2 \mathfrak{F}( 4 \lambda^2 \Delta_T(t) ),
\end{equation}
where the maximization over $t$ is carried out using a Nelder–Mead algorithm, while the function $\mathfrak{F}( 4 \lambda^2 \Delta_T(t) )$ is evaluated with a semidefinite program for each $t$.

\begin{figure}[th]
  \includegraphics{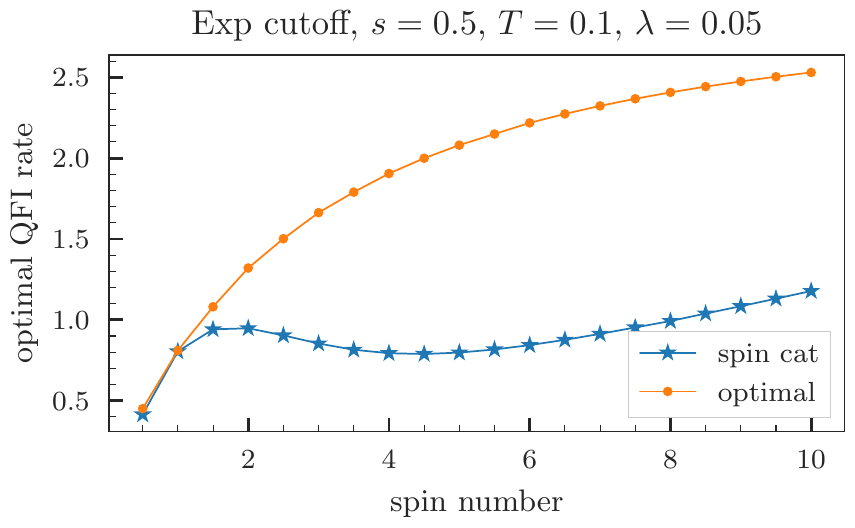}
  \includegraphics{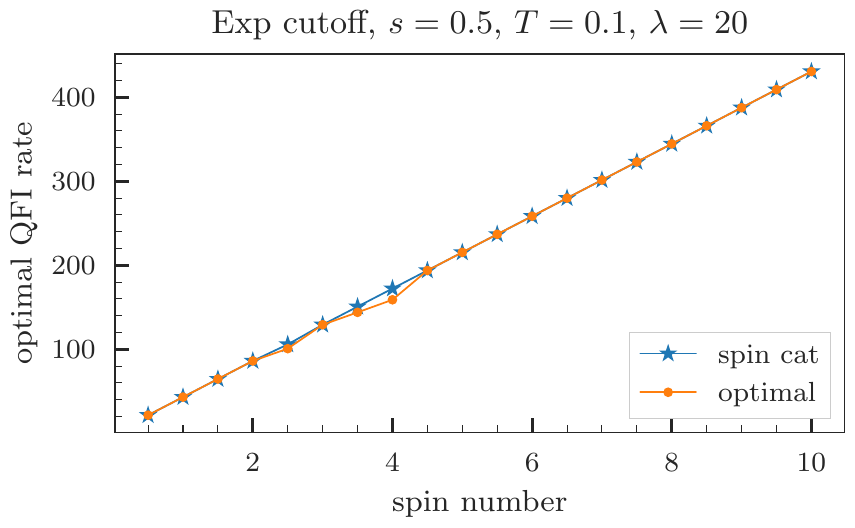}
  \includegraphics{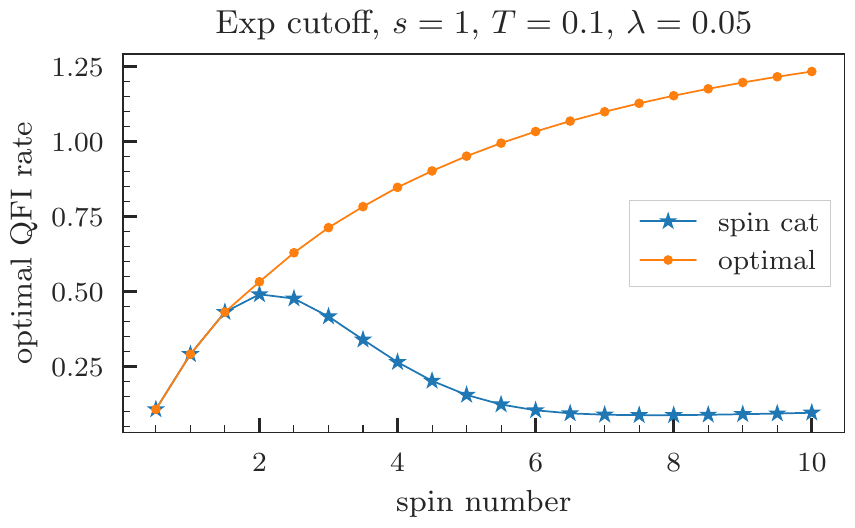}
  \includegraphics{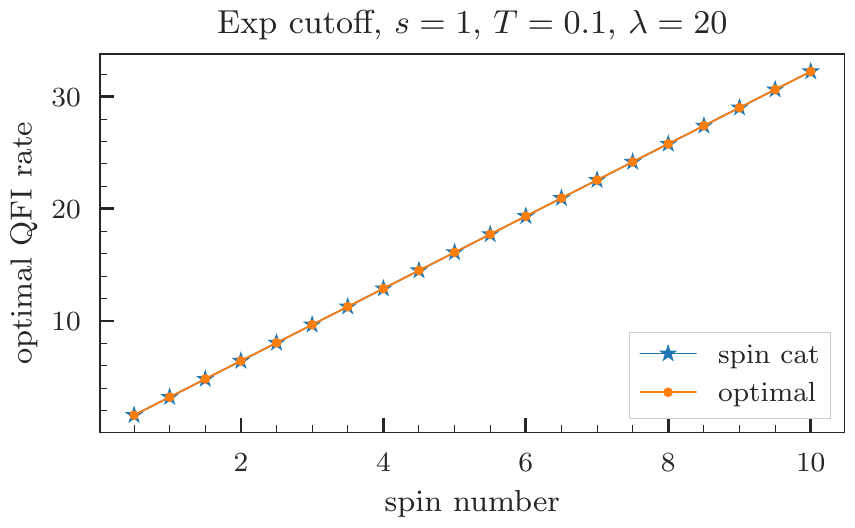}
\caption{Comparison between dephasing thermometry with optimal spin probes and with GHZ states.}
\label{fig:GHZvsOptimal}
\end{figure}

Some of the results of this comparison are shown in Fig.~\ref{fig:GHZvsOptimal}.
We see that for both $s=1$ and $s=0.5$, the performance of spin cat states coincides with the optimal result for very strong coupling (e.g. $\lambda=20$ in the two right panels of Fig.~\ref{fig:GHZvsOptimal}).
However, in the weak coupling regime (e.g. $\lambda=0.05$ in the two left panels) the spin cat states starts as optimal, then become suboptimal with the optimal QFI rate that decreases as the spin number $j$ increases, but after this decline the optimal QFI rate starts increasing again with $j$.
Unfortunately, evaluating the quantity in Eq.~\eqref{eq:numerical_topt} for larger values of $j$ is too computationally demanding.
However, we suspect that eventually, for $j$ large enough, spin cat probes may become optimal again, since for this class of states increasing $j$ is equivalent to increasing $\lambda$ and for large $\lambda$ we have shown that they are optimal.
Similar conclusions may be found for different values of $s$ and $T$.

A previous indication of the optimality of spin cat probes for the estimation of environmental parameters appearing in the dephasing factor was given in~\cite{Rossi2015}\footnote{More precisely Greenberger–Horne–Zeilinger states of multi-qubit systems were considered instead of cat states of a spin-$j$ system, but the analysis is equivalent.}.
In particular, it was shown that for fixed $j$ there is a threshold value, if the dephasing is weak enough it is optimal to use spin cat states.
We note, however, that the figure of merit optimized in~\cite{Rossi2015} was the QFI and not the QFI rate.
Moreover, for a \emph{fixed} dephasing $\Delta$, i.e. the quantum channel~\eqref{eq:dephasing_est_Appendix}, the optimal state for asympotitcally large $j$ is not a spin cat state~\cite{Knysh2013}.
However, for a time-dependent problem the scenario is quite different, since the optimal time also scales with $j$ and the problem differs from the estimation a fixed dephasing factor $\Delta$.

\section{Short-time expansions}
\label{app:short_time}
We report here the short-time expansions of absorbed heat, dephasing function and QFI for the spectral density with exponential cutoff. 
For conciseness we express $t$ and $T$ in units of $\omega_c$.
The heat reads:
\begin{align}
  \mathcal{Q}(t) &= \lambda^2 \left[ q^{(2)} t^2 + q^{(4)} t^4 + O(t^6) \right] \\
   q^{(2)} &= s (s+1) \Gamma (s) \\
   q^{(4)} & = -\frac{1}{12} \left(s^4 + 6 s^3+11 s^2 +6 s \right) \Gamma (s),
\end{align}
the dephasing function:
\begin{align}
  \Delta_T(t) &= \Delta_T^{(2)} t^2 + \Delta_T^{(4)} t^4 + O(t^6)\\
  \Delta_T^{(2)} &= \frac{2 (s-1) s \Gamma (s-1) \left[ 2 (s-1) s T^{s+1} \Gamma (s-1) \zeta (s+1,T+1)+\Gamma (s+1)\right]}{\Gamma (s+1)} \\
  \Delta_T^{(4)} &= -\frac{\left[s \left(s^3+2 s^2-s-2\right) \Gamma (s-1) \left(2 (s-1) s T^{s+3} \Gamma (s-1) \zeta (s+3,T+1)+\Gamma (s+1)\right)\right]}{6 \Gamma (s+1)}
\end{align}
from which the QFI reads:
\begin{align}
  \mathcal{F}[\rho_T(t)] = \frac{ 16 \lambda^4 \, \left[ \partial_T \! \Delta_T(t)\right]^2 }{\exp\left[ 8 \lambda^2 \Delta_T(t) \right] -1 } &= f^{(2)} t^2 + f^{(4)} t^4 + O(t^6) \\
  f^{(2)} &= \frac{\lambda^2 \left( \partial_T \! \Delta_T^{(2)}\right)^2 }{2 \Delta_T^{(2)}} \\
  f^{(4)} &= -\frac{\lambda^4 (\Delta_T^{(2)})^2 (\partial_T \!\Delta_T^{(2)})^2 + \lambda^2 \left[ \Delta_T^{(4)} (\partial_T \! \Delta_T^{(2)})^2 - 2 \Delta_T^{(2)} \partial_T \Delta_T^{(2)} \partial_T \Delta_T^{(4)} \right]}{2 (\Delta_T^{(2)})^2}.
\end{align}
As long as the coefficient $f^{(4)}$ is negative (this depends on the particular parameter values, but we can always find  $\lambda$ large enough for which this holds) the optimal time is
\begin{equation}
  t_\mathrm{opt} = \mathrm{argmax}_{t} \frac{ \mathcal{F}[\rho_T(t)]}{t} = \sqrt{- \frac{f^{(2)}}{ 3 f^{(4)}} },
\end{equation}
which approaches zero as $\lambda^{-1}$ for $\lambda \to \infty$.

\end{document}